\documentstyle[12pt,aasms4]{article}

\begin{document}
\title{ Evidence for Old Stars in the  \\
Red Low Surface Brightness Galaxies \\ UGC~6614 and F568-6}

\author{
A.~C.\ Quillen\altaffilmark{1}$^,$\altaffilmark{2} \&
T.~E.\ Pickering\altaffilmark{1}$^,$\altaffilmark{3} 
}
\altaffiltext{1}{University of Arizona, Steward Observatory, Tucson, AZ 85721}
\altaffiltext{2}{E-mail: aquillen@as.arizona.edu}
\altaffiltext{3}{E-mail: tim@as.arizona.edu}

\def\spose#1{\hbox to 0pt{#1\hss}}
\def\lta{\mathrel{\spose{\lower 3pt\hbox{$\mathchar"218$}}
     \raise 2.0pt\hbox{$\mathchar"13C$}}}
\def\gta{\mathrel{\spose{\lower 3pt\hbox{$\mathchar"218$}}
     \raise 2.0pt\hbox{$\mathchar"13E$}}}

\begin{abstract}
We present near-Infrared images of low surface
brightness galaxies: $H$ band images of UGC~6614 and F568-6 (Malin 2).  
The $H$ band images show spiral structure so we are
confident that we are seeing the galaxy disks as well as
the central bulges.
The optical-IR colors of these galaxies, $R-H =  2.2\pm 0.2$, 
$B-H  = 3.5-4.2$, are 
extremely red and are similar to those of many S0 and elliptical galaxies.
This represents strong evidence for a component of old stars.
Recent studies find that low surface brightness galaxies 
span a wide range of morphologies and colors.  
In this context UGC~6614 and F568-6 could be examples of 
low surface brightness galaxies which have undergone a past epoch 
of more vigorous star formation.
\end{abstract}


\section {Introduction}

Most low surface brightness galaxies (LSBGs) discovered in photographically
based surveys are unusually blue despite the lack of significant 
ongoing star formation, 
with mean colors $B-V = 0.49 \pm 0.04$  (\cite{mcg94}), and
$B-V =  0.7\pm 0.05$ for the large scale length sample of \cite{spray95}.
Generally the stellar populations of LSBGs are 
metal poor ($Z \sim 1/3$ solar, \cite{mcg94}, also \cite{vdh93} and \cite{sch90}).  
Because of their blue colors and low surface brightness
they are expected to be difficult to detect in the near-IR.
In fact few LSBGs have been observed in the near-IR, with perhaps the only
examples presented in \cite{kne94}.
For recent reviews on LSBGs see \cite{imp97} or \cite{bot97}.

However, in our recent work studying the spiral structure of
the large scale length LSBGs UGC~6614, 
and F568-6 (otherwise known as Malin 2),
(\cite{qui97}) we found that the disks of these galaxies
probably contain significant stellar mass surface densities
(greater than a few $\times 10^{10} M_\odot$) and so have processed
into stars a gas mass greater than that now observed in \ion{H}{1}.
These galaxies also have quite red colors and
high metallicities (near or greater than solar (\cite{pic97b}; \cite{mcg94})
compared to most LSBGs suggesting an old stellar population which would have 
moderate mass-to-light ratios consistent with our mass estimates.  
This prompted us to consider the possibility that this postulated older 
stellar population component would be easily detectable in the near-IR.

In this paper we report detections of low surface brightness
galaxies in the near-IR bands.   We emphasize here that the two low surface
brightness galaxies observed here were chosen precisely because of their
red colors, high metallicities and prominent spiral arm morphology
which makes them 
exceptional compared to many currently cataloged LSBGs but 
similar to normal high surface brightness galaxies. 
We note that recent studies (\cite{kne93} and \cite{one97b})
have emphasized that LSBGs have a wide variety of morphological and color
properties.

\section {Observations}


The $H$ images were obtained
on the $61$'' Telescope of Steward Observatory  on Mt. Bigelow
on 1997 Feb 22 and 23 with a $256 \times 256$ NICMOS3 infrared array 
with a spatial scale of $0.90$ arcsec/pixel.
Individual images were taken with an exposure time of $30$ seconds 
alternating between object and sky. 
Total on source integration times were $45$ minutes with 
equivalent time spent on the sky.
Flat fields were
constructed from median filtered sky frames.  Images were aligned to the
nearest pixel and combined to form the final images.
The $H$ band images were observed during photometric conditions and were
calibrated on the CTIO/CIT system using standard stars listed by
\cite{car95}.

The $B$ and $V$ images were obtained
on the $61$'' Telescope of Steward Observatory on Mt. Bigelow
on 1997 Apr. 8 with a 1kCCD  binned $2\times 2$ so the 
resulting spatial scale is $0.40$ arcsec/pixel.
Individual images were taken with an exposure time of $300$ seconds
and total integration times were $25$ minutes in both bands.
These images were observed during photometric conditions and
were calibrated to the Johnson system using three standard fields from 
\cite{lan92}.
The $R$ band images were taken from \cite{pic97} and are also
displayed in \cite{qui97}.

For UGC~6614, our values for $B-V$ which range between 
1.0 and 0.5 at larger radius, agree within 0.05 mag 
with those of \cite{mcg94b}.
The $R$ band surface brightness radial profile of UGG 6614 agrees 
within 0.2 mag with that of \cite{kne93} and that of \cite{deb95}.


\section {Results}
In Fig. 1 are displayed grayscale $B,R$ and $H$ images of the two galaxies.
Spiral structure seen in the optical images
is also clearly evident in the $H$ band images.  
We therefore detect in $H$ band emission from the disks of these galaxies,
as well as emission from their bulges.
For both galaxies the spiral structure is more well defined in the $B$ band
than in the red $R$ or $H$ bands so that  
the spiral features become smoother with increasing wavelength.
In normal high surface brightness galaxies spiral arms are also typically
sharper in the bluer bands and smoother in the near-IR or red bands.
The dependence of morphology on wavelength is a result of 
absorption from dust and emission from bright young blue stars
both which are more prominent in bluer bands and are 
concentrated in narrow features concentrated along the spiral arms
(see for example \cite{gon96}).  It is likely that the same is true in 
the two low surface brightness galaxies observed here.

In Fig. 2 we show radial surface brightness profiles in all bands
observed.
We note that in UGC~6614 the ring at $r\sim 30''$ is more sharply defined
in the bluer bands.  
The ring which is evident as a peak in 
the radial surface brightness profile in $B$ band 
could be more accurately described as a plateau in $H$ band.

Fig. 2 also plots the various colors as a function of radius 
for both galaxies.  We note that colors in all bands are quite quite red, 
though
the ring in UGC~6614 at $r \sim 30''$ is bluer than the surrounding disk.
Both galaxies show evidence for a color gradient in
the sense that they become bluer with increasing radius.
This type of color gradient is also observed in normal or high surface 
brightness 
galaxies and is consistent with an age or/and metallicity gradient
(e.g. \cite{dej96}).

In Fig. 3 we show the colors of UGC~6614 and F568-6 
in a color/color plot compared to those of other galaxies and clusters.
The nucleus of both galaxies has red colors typical of a normal galaxy bulge
which is similar to that of many elliptical galaxies.  
The outer disk of both galaxies, with $B-V \sim 0.5$ and $B-H \sim 3.8$,
has colors similar to some Sc nuclei and older LMC or SMC clusters.
It is likely that at large radius the disks are either younger or/and less 
metal rich, as is true in normal galaxies (e.g. \cite{fro88}; \cite{dej96}).
We see that a significant fraction of the stellar population in these
two LSBGs cannot
be from young and metal-poor stars because the colors are too red, both
in the bulge and in the outermost disk points measured.

\section {Discussion}

In this paper we have presented $B,V,R$ and $H$ band images of two
low surface brightness galaxies UGC~6614 and F568-6 (Malin 2).  
Because we see spiral structure in our $H$ band images we are confident that 
we detect in $H$ band emission from the disks of these galaxies,
as well as emission from their bulges.
Because of the blue colors 
of the majority of currently cataloged low surface brightness galaxies
and their low surface brightnesses, 
they were expected to be undetectable in the near-IR.  
We were only able to detect UGC~6614 and F568-6 at 
$H$ band because they are extremely red.

In fact we have found that these galaxies 
have extremely red optical to near-IR colors, with 
$R-H = 2.2 \pm 0.2$ and $B-H  = 3.5 - 4.2$.
We find that these colors are similar to those of many S0 and elliptical 
galaxies and and so are strong evidence for 
the presence of an old stellar population.  
As noted in previous studies (\cite{fro85}) the nuclei of 
many high surface brightness spiral
(Sc) galaxies have similar red colors.

We find that the spiral arms of these two galaxies are more prominent 
in the $B$  band than in $R$ or $H$ band.
As has been found in normal spiral galaxies, 
the spiral arms of UGC~6614 and F568-6 become smoother with increasing 
wavelength.  In our previous work, \cite{qui97}, by considering
the strength of the spiral structure, we found that the disk of the galaxy
had a significant stellar mass compared to the atomic gas component 
(stellar/gas mass $\gta 1$).  Since the disk mass limit depends on
strength of the spiral structure, the smoother appearance of the
$H$ band images suggests that the disks of these galaxies
are somewhat more massive than we estimated in \cite{qui97}.

The large estimated stellar disk mass in UGC~6614 and F568-6 suggests
that a significant gas fraction has been converted into stars.
This would be consistent with the near solar metallicities
measured by \cite{pic97b} and \cite{mcg94}.
Since \cite{pic97b} also find that the current level
of star formation (based on the number of HII regions detected)
is low,   if UGC~6614 and F568-6 have a substantial older stellar
population as suggested by their red colors and the disk mass limits,
then it is likely that these galaxies have undergone
a previous epoch of more vigorous star formation.
This suggests that the evolutionary history of these two LSBGs is not unsimilar
to many nearby normal or high surface brightness spiral galaxies.

The possibility of an older stellar population in UGC~6614 and F568-6
would make these LSBGs quite different than the majority which have blue
colors, low metallicities and are suspected to have young stellar 
populations.  
As summarized by \cite{mcg94b}, for the majority of LSBGs,
despite the ``absence of significant star formation,
the most plausible scenario is a stellar population with a young mean age
stemming from late formation and subsequent slow evolution.  These properties
suggest that LSBG disks formed from low initial overdensities with 
correspondingly late collapse times.''  
We note, however, that with a novel IMF, \cite{pad97} has succeeded in matching
the extremely blue colors of some LSBGs with an old (10 Gyr) 
stellar population.  The fact that LSBG galaxies
exist with a range of colors (suggesting a variety
of evolutionary scenarios) represents a substantial challenge
for cosmological galaxy formation studies (e.g. \cite{dal97}) as well as
in understanding the process of star formation at low surface densities.

A recent CCD based multicolor survey has discovered that 20\% of
recently identified LSBGs have $B-V > 0.9$ (O'Neil et al. 1997a,b).
If these red LSBG galaxies are similar in stellar population to UGC~6614 and 
F568-6 then they could be galaxies
which may have had previous epochs of star formation.
One intriguing possibility is that 
they may have been part of the faint blue galaxy population.




\acknowledgments

We thank M. Rieke for making these observations possible by repairing
the camera during the observing run.
We also acknowledge helpful discussions and correspondence with 
P. Knezek, J. Frogel, C. Impey, G. Bothun, K. O'Neil, and H.-W. Rix.
We thank Ray Bertram for obtaining images of Malin 2.
We acknowledge support from NSF grant AST-9529190 to M. and G. Rieke.

\clearpage


\newpage
\clearpage

\begin{figure*}
\caption[junk]{Greyscale $B$, $R$ and $H$ band images of UGC 6614 and F568-6.
\label{fig:fig1} }
\end{figure*}

\begin{figure*}
\caption[junk]{$B,V,R$ and $H$ band surface brightnesses and colors as a 
function of semi-major axis length (or radius from the nucleus) 
for UGC 6614 and F568-6.  
In the bottom plots $B,V,R$ and $H$ bands are
shown as hexagons, pentagons, squares, and triangles respectively.
Note that the ring at $r\sim 40''$ in UGC 6614 is more sharply
defined in the blue optical bands.
Both galaxies become bluer with increasing radius.
Fluctuations in the sky are observed in the $H$ band images at the level
of 22.2 mag/''$^2$ so that points at surface brightnesses fainter than this
are unreliable.  In $B,V$ and $R$, fluctuations
in the sky (due to errors in the flat fields and reflected light) 
dominate at $r \gta 70''$ for both galaxies.
\label{fig:fig2} }
\end{figure*}

\begin{figure*}
\caption[junk]{ 
Comparison of bulge and disk colors in UGC~6614 and F568-6
to those of other galaxies and globular clusters.
Colors for UGC~6614 and F568-6 are shown for a range of radius.
Since the colors of these tow galaxies are so similar only one set
of points is shown for both.
The reddest colors correspond to the bulge and the bluest colors
to the outermost disk (see Fig. 2).
Colors have been corrected for galactic extinction (represented
by the zero subscript in the axis labels).
The colors in UGC 6614 range between colors typical
for Sc galaxies to those of SO and Elliptical galaxies.
The data for the LMC and SMC globular clusters are from Persson et al. (1983).
The data for the E and SO galaxies are from 
Persson, Frogel \& Aaronson (1979).  
The data for the Sc galaxies are from Frogel (1985).
The data for the M31 globular clusters are from Frogel, Persson \& Cohen (1980).
\label{fig:fig3} }
\end{figure*}

\end{document}